# A Review of Papers that have a bearing on An Analysis of User Interactions in A Collaborative On-line Laboratory

Vita Hinze-Hoare

December 2006

School of Electronics and Computer Science
Southampton University
Southampton
SO17 1BJ

**Abstract**

A number of papers have been reviewed in the areas of HCI, CSCW, CSCL. These have been analyzed with a view to extract the ideas relevant to a consideration of user interactions in a collaborative on line laboratory which is being under development for use in the ITO BSc course at Southampton University.

The construction of new theoretical models is to be based upon principles of collaborative HCI design and constructivist and situational educational theory.

An investigation of the review papers it is hoped will lead towards a methodology/framework that can be used as guidance for collaborative learning systems and these will need to be developed alongside the requirements as they change during the development cycles. The primary outcome will be the analysis and re-design of the online e-learning laboratory together with a measure of its efficacy in the learning process.

CONTENT

*"ICT supported learning is not an objective in itself but is indispensable for bringing about the socio economic changes in which the European Union has engaged itself".*

*European ODL Liaison Committee 2004*

# 1. INTRODUCTION

The twin fields of Computer Supported Collaborative Work (CSCW) and Computer supported Collaborative Learning (CSCL) have been the subject of intense interest in the HCI research community during the past seven years.

The split between CSCW and CSCL has grown wider in response to the recognition that the learning process is more distinct from the working pattern and is more intensively understood through new theories of pedagogy and education.

It has become apparent that CSCL requires all of the facets of CSCW but in addition is constrained by these pedagogical theories and as such it is argued here that CSCL is a subset of CSCW (see figure1)

The process of research is also a learning process but one which is more highly refined and involves learning in a particular way using special techniques and tools. As such it is argued further that research which is supported by computer collaboration is a subset of CSCL, see Figure 1.

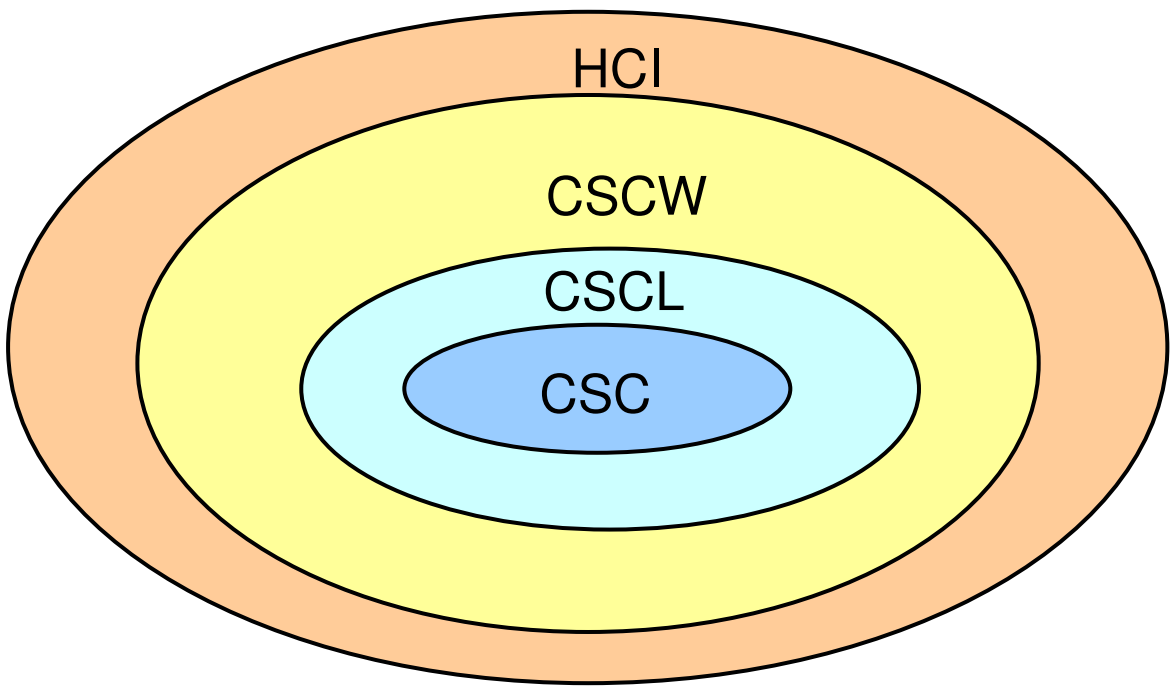

**Figure 1**

*The Layered model of collaboration within HCI*

# 2. Human Computer Interaction (HCI)

## 2.1. History of HCI

According to Diaper (2005) the chronology of HCI starts in 1959 with Shakel's paper on "The ergonomics of a computer" which was the first time that these issues were ever addressed. This was followed by Licklider who produced what has come to be known as the seminal paper (1960) on "Man – Computer Symbiosis" which sees man and computer living together. There was no further significant activity for almost 10 years when in 1969 the first HCI conference and first specialist journal, "The International Journal of Man-Machine Studies" was launched. The 1980s saw the launch of three more HCI journals and conferences with an average attendance of 500 (Diaper 2005). It was not until the 1990s that the "I" in HCI switched from "interface to "interaction" reflecting the vastly expanding range of digital technologies. It was also during the 1990s that the term "Usability" has come to be synonymous with virtually all activities in HCI. Prior to this HCI encompassed five goals to develop or improve:

- Safety
- Utility
- Effectiveness
- Efficiency
- Usability

Originally usability was the least but has since been promoted to cover everything. "The study of HCI became the study of Usability" (Diaper, 2005).

## 2.2. The Basic Characteristics and Structure of HCI

HCI has become an umbrella term for a number of disciplines including theories of education, psychology, collaboration as well as efficiency and ergonomics as shown in Figure 2: HCI umbrellaFigure 2 Figure 1below.

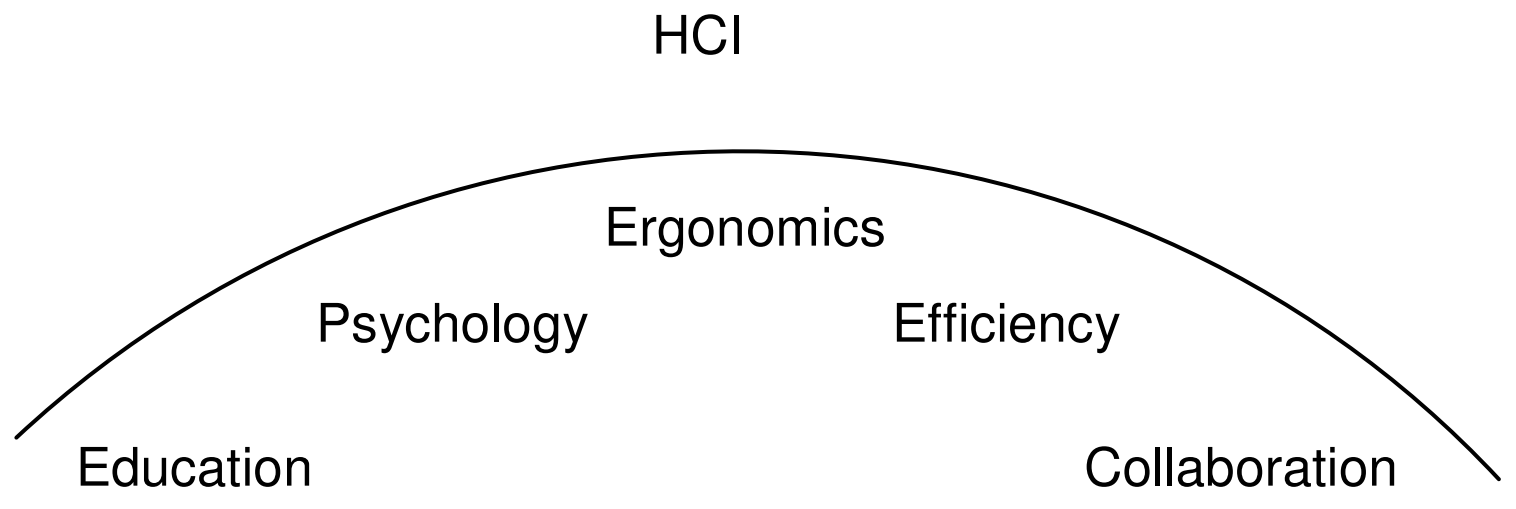

**Figure 2: HCI umbrella**

Recent developments in the area of HCI have shown an interest in adaptive interfaces, speech recognition, gestures and the role of time.

#### 2.2.1. Adaptive interfaces

Savidis and Staphanidis (2004) discuss the key development requirements for building universal accessibility into Interfaces. By universal access they mean that anyone can interact with any computer at anytime in any place for virtually any task. They conclude that it is important that interface elements reflect the design of well known metaphors such as the desktop metaphor.

One way of providing universal access is to apply user interface adaptation which is a dynamic interface behaviour that changes the interface on the fly according to the tasks which are being performed to make it easier for the user. However, this requires a degree of programming sophistication. For example an adaptive interface might present the user with three or four windows and when the user chooses one to work in this could automatically expand in size at the expense of the others automatically adapting to the users' needs. This might be exploited further by the inclusion of built in cameras to track the users' eye movements and respond adaptively to what the user is looking at.

#### 2.2.2. Speech Recognition

Wald (2005) has studied the role of automatic speech recognition (ASR) to record lecturer's notes. Standard ASR software produced little more than stream of unpunctuated text. A new application ViaScribe has been developed in collaboration with IBM to format the ASR transcription and to provide a readable real time display and an archived transcription service. He concludes that synchronised ASR can make some text and visual material more accessible to deaf students.

#### 2.2.3. Temporal issues in HCI

Wild and Johnson (2004) look at the role of time with respect to HCI. They draw a distinction between m-time (monochronic or "one thing at a time" approach based on using schedules) and p-time (polychronic – multitasking time) and refer to six dimensions for the analysis of temporal issues

- Duration
- Temporal location ( over a period of time)
- Sequence
- Deadline
- Cycle ( regular occurrence)
- Rhythm (variation of intensity)

Wild and Johnson conclude that these need to be included in any HCI analysis.

Oulasvirta and Tamminen (2004) discuss the role of temporal tension, by which they mean time pressure to meet deadlines and how it affects collaborative working. They identify five effects of time and deadlines on the speed and rate of work.

- Acceleration
- Deceleration
- Hurrying
- Normal or balanced
- Waiting

They conclude that time is an under researched topic in HCI and needs greater investigation. Temporal tension is work driven rather than device driven and this might have some application in considering the design of the interface for the collaborative e lab.

## 2.3. HCI Theories and Principles

There are typically many thousands of rules which have been developed for the assessment of usability (Nielsen, J. 1993, p19), and there have been many attempts to reduce the complexity to a manageable set of rules (Nielsen, J. 1993, Baker, Greenberg and Gutwin, 2002 and Hinze-Hoare, 2004).

Jacob Nielsen has produced 10 rules which he calls usability heuristics and which are designed to explain a large proportion of problems observed in interface design. Nielsen recommends that these principles should be followed by all user interface designers and they are as follows:

- **Simple and natural dialogue**
  Efforts should be made to avoid irrelevant information. Nielsen says that every extra unit of information competes with units of relevant information and diminishes its visibility.

- **Speak the Users' language**
  All information should be expressed in concepts which are familiar to the user rather than familiar to the operator or the system.
- **Minimize the Users' memory load**
  It is important that the user should not have to remember information from one part of a dialogue to another. Help should be available at easily retrievable points in the system.
- **Consistency**
  Words situations and actions should always mean the same thing no matter where they occur in the system.
- **Feedback**
  Users should always be informed about what is going on in the system in a timely and relevant way.
- **Clearly marked Exits**
  Errors are often made in choosing functions which are not required and there needs to be a quick emergency exit to return to the previous state without having to engage in extended dialogue.
- **Shortcuts**
  Required by the expert user (and unseen by the novice user) to speed the interaction with the system.
- **Good error messages**
  These need to be expressed in a plain language that the user understands which are specific enough to identify the problem and suggest a solution.
- **Prevent Errors**
  A careful design will prevent a problem from occurring.
- **Help and documentation**
  the best systems can be used without documentation. However, when such help is needed it should be easily to hand, focused on the users task and list specific steps to solutions.

(From Jakob Nielsen, 1993, Usability Engineering, Chap 5, Usability Heuristics.)

Baker, Greenberg and Gutwin (2002) have taken Jakob Nielsen's heuristic evaluation a stage further and considered the problems posed by groupware usability concerns. They have adapted Nielsen's heuristic evaluation methodology to collaborative work within small scale interactions between group members. They have produced what they call 8 groupware heuristics.

- **Provide the means for intentional and appropriate verbal communication.**
  The most basic form of communication in groups is verbal conversation. Intentional communication is used to establish common understanding of the task at hand and this occurs in one of three ways.
- **People talk explicitly about what they are doing**
- **People overhear others conversations**
- **People listen to running commentary that people produce describing their actions.**
- **Provide the means for intentional and appropriate gestured communication.**
  Explicit gestures are use alongside verbal communication to convey information. Intentional gestures take various forms. Illustration is acted out speech, Emblems are actions that replace words and Deixis is a combination of gestures and voice
- **Provide consequential communication of an individual's embodiment**
  Bodily actions unintentionally give off information about who is in the workspace, where they are and what they are doing. Unintentional body language is fundamental for sustaining teamwork.
- **Provide consequential communication of share artifacts**
  A person manipulating an artifact in a workspace unintentionally gives information about how it is to be used and what is being done with it

- **Provide Protection**
  People should be protected from inadvertently interfering with the work of others or altering or destroying work that others have done
- **Manage the transitions between tightly and loosely coupled collaboration**
  Coupling is the degree to which people are working together. People continually shift back and forth between loosely and tightly coupled collaboration as they move between individual and group work
- **Support people with the coordination of their actions**
  Members of a group mediate their interactions by taking turns negotiating the sharing of a common workspace. Groups regularly reorganize the division of work based upon what other participants have done or are doing.
- **Facilitate finding collaborators and establishing contact**
  Meetings are normally facilitated by physical proximity and can be un scheduled, spontaneous or initiated. The lack of physical proximity in virtual environments requires other mechanisms to compensate.

(From Baker, Greenberg and Gutwin 2002)

Hinze-Hoare (2004) has performed an analysis of HCI rules based upon citation frequency of authors. The top eight rules were found to be:

**1. Familiarity**

This is the degree to which the user's own real world personal experience and knowledge can be drawn upon to provide an insight into the workings of the new system. The familiarity of a user is a measure of the correlation between their existing knowledge and the knowledge required to operate the new system. To a large extend familiarity has its first impact with the users' initial impression of the system and the way it is first perceived and whether the user can therefore determine operational methods from his own prior experience. If this is possible this greatly cuts down the learning time and the amount of new knowledge that needs to be gained. The term familiarity is proposed by Dix *et al* (1992) but is referred to by other authors under different terms for example Jordan *et al* (1991)[12] refers to familiarity as guessability. Schneiderman (1998) and Preece (1994) each refer to familiarity in terms of the reduction of cognitive load.

**2. Consistency**

Consistency, according to Dix *et al* (1992) relates to the likeness in behavior arising from similar situations or similar task objectives. He also thinks that this is probably the most widely mentioned principle in the literature on user interface design. It is considered of vital importance that the user has a consistent interface.

**3. Forward Error Recovery**

This is the ability of users recovering from their errors, which they invariably make. There are two directions in which recovery can occur both forward and backward. Forward error recovery involves the prevention of errors. Backward error recovery concerns the easy reversal of erroneous actions. The latter is usually concerned with the users actions and is initiated by the user. The former is one, which should be engineered into the system and initiated by the system. In this sense recoverability is connected to fault tolerance, reliability and dependability. Ken Maxwell (2001) considers this basic usability a level one priority, which he calls error protection. Jeff Raskin (2000) rates this as part of his first law of interface design, which states, "a computer shall not harm your work or through inaction allow your work to come to harm".

**4. Substitutivity**

This concerns the ability of the user to enter the same value, or perform the same action in different ways according to his own personal preference. For example a user might wish to enter values in either inches or centimeters, or he may wish to open a program with the mouse or with the keyboard. This input Substitutivity contributes towards an overall flexible HCI structure, which allows the user to choose whichever he considers most suitable. Schneiderman (1998) and Preece (1994) provide a specific example of providing shortcuts as an alternative.

**5. Dialogue Initiative**

This allows the user the freedom from artificial constraints on the input dialogue boxes. When humans communicate with computers and the dialogue is set up it is important to ensure that the human partner has the initiative in the conversation. If the system initiate all dialogues and the user simply responds, this is called “system preemptive”. On the other hand if the user is free to initiate actions then this is called “user preemptive”. An effective system is one, which maximizes the users ability to preempt the system, and minimizes the systems ability to preempt the user. This is also referred to as allowing input flexibility, Preece (1994)

**6. Task Migratablility**

This concerns the transfer of control for executing tasks between the system and the user. Checking the spelling of a document is a good example. The user can quite easily check the spelling for himself by the use of a dictionary. However the task is made considerably easier if it can be passed over to the system to perform with simple checks made by the user as to the acceptable spelling i.e. the difference between US and British Dictionaries. This is an ideal task for automation. However, it is not desirable to leave it entirely in the hands of the computer as dictionaries are limited and therefore the task needs to be handed over to the user at those complex points where the system cannot cope. Ken Maxwell (2001) talks of this as level two collaborative organisational interaction which he considers being a high level of HCI interaction. This is the ability of the interface to hand the task over to the user so that the initiative rests with the human side of the interaction. This can be measured by the degree of performance available through the use of unfamiliar tasks.

**7. Responsiveness**

This measures the rate of communication between the user and the system. It covers such areas as simple response time to keys pressed or to actions executed but also concerns the quality of the feedback provided by the system and the appropriateness of the responses that are made. For instance a computer system that takes 10 seconds to respond to a simple click on a toolbar button will rapidly lead to user frustration. On the other hand a system that provides rapid and appropriate responses, including both audio and visual cues will be well received by the user. Jeff Raskin (2000) writes of this as his second law of the computer interface which he defines as “a computer shall not waste your time...”. This is the immediacy and feel that the interface provides to the user and is an important subjective criterion of perceived efficiency.

**8. Customisability**

This is ability of the user to modify the interface. This is sometimes known as adaptability and allows different users to adapt the interface according to their own level and style of interaction. This customisation can be enhanced by providing the user with various levels of interaction. At the lowest end it may merely involve the altering of colours and styles while at the upper end an application may allow the user to develop their own macros or even write their own VB modules. This allows for the elimination of repetitive tasks or for the construction of a personally more suitable environment to the user.

## 2.4. HCI Models

### 2.4.1. Audience participation model

Nemirovsky (2003) presents a new theoretical model for audience participation in HCI. He considers that the old perspective is that of computers as deterministic boxes blindly following their commands while users are incapable of changing the course of the program running on the computer. To this he presents an alternative and proposes that users should be considered as an audience rather than participants. Old models include the idea that the mass of people wish to be entertained rather than to be creative and are punished for any creative thinking while using a computer which is regarded as making a mistake. As a consequence computer users do not have a proper framework to express themselves. This is a strikingly radical approach. Instead Nemirovsky is concerned with users as an audience that explore the media space. He goes on to discuss the emonic environment which he defines as a framework for creation, modification, exchange and performance of audio visual media. This is composed of the three layers:

- Input (interfaces for sampling)

- Structural (a neural network for providing structural control)
- Perceptual (direct media modification)

This is unlikely to have any real application to the CSCL problem at hand.

## 2.5. HCI Analysis Methodologies

A number of different methodologies have been created to determine the effectiveness of HCI measurements. More will be said about this in the sections on CSCW and CSCL however one is noted here.

### 2.5.1. User Needs Analysis

Lindgaard *et al* (2006) present a user needs analysis methodology which suggests how and where user centred design and requirements engineering approaches should be integrated. After reviewing various process models for user centred design analysis they suggest their own approach.

Lindgaard *et al* first identify the main problems as follows

- The decision where to begin and end the analysis is never clear
- Deciding how to document and present the outcome
- The process involves the following steps:
    - First: Identify user groups and interview key players from all groups to find the different roles and tasks of the primary and secondary users
    - Second: Communicate this information to the rest of the team by constructing task analysis data and translating this into workflow diagrams supporting the user interface design. Create a table that shows the information about user roles and data input
    - Third: Upon submitting the first draft of the user needs analysis report create the first iterative design prototype of the user interface based on minimising the path of data flow. (Initially prototyping in PowerPoint was faster and more effective that prototyping in Dreamweaver).
    - Fourth: Prototypes were handed over to developers as part of the user interface specification package.
    - Fifth: Usability testing was used to determine the adequacy of the interface. Feedback from watching users work with the prototype and discussing with them what they were doing always resulted in more information.
    - Sixth: Prototype usability testing meant that the requirements became clearer which resulted in more changes to the user interface design and the prototype.
    - Seventh: The formal plan involved three iterations of design- prototype- usability test for each user role (they could not keep to this and had no more that two test iterations and in most cases only one)
    - Eighth: Practical issues of feasibility should not be overlooked in the quest to meet users' needs. A highly experienced software developer is a necessity on the user interface design team in order to ensure that the changes were proposed were feasible ( in some cases interface ideas were dropped because they were not feasible, take too long or cost too much).

## 2.6. Problems of HCI

### 2.6.1. The Fragmentation of HCI

The History of HCI according to Diaper (2005) shows a lack of coherent development. There is no agreement as to

- What HCI should be

- What HCI can do
- How HCI can do it
- How HCI can be allowed to do it

Proposals that failed to address these issues are likely to lack substance and need to be addressed. The discipline is becoming increasingly fragmented to the point where it is difficult to establish consensus in the field.

The fragmentation of discipline of HCI is already so extensive according to Diaper 2005 that it is hard to even characterise the method of approach. As an example different practitioners have different priorities and different methodologies. Some approaches will start with requirements, while others start with evaluation and yet others with dialogue or user modelling or scenarios or information or design or artefacts or processes. This lack of agreement highlights the necessity for the development of a general systems model, both in the general HCI approach and in the specific collaborative approach.

Furthermore, Kligyte, (2001) has recorded that “CSCL emerged as an autonomous research field out of the wider CSCW research area quite recently, and there is still lack of consensus about core concepts, methodologies and even the object of study”.

Lipponen (2002) concludes his review paper on CSCL with the comment “There is still no unifying and established theoretical framework, no agreed objects of study, no methodological consensus or agreement about the concept of collaboration or unit of analysis.

Strijbos *et al* (2005) have pointed out that a review of CSCL conference proceedings revealed a general vagueness in the definitions of units used in the analysis of CSCL. They further comment that arguments were lacking in choosing units of analysis and reasons for developing content analysis procedures were not made explicit. They conclude that CSCL is still an emerging paradigm in educational research.

## 2.7. HCI summary

We have seen that HCI theories are not yet fully established and that the discipline is highly fragmented making it difficult to characterise a single method of approach or even an accepted method. The lack of agreement between authorities in this field suggests that the approach must be carefully tailored to the specific needs of the environment to which they are applicable.

# 3. Computer Supported Collaborative Working (CSCW)

## 3.1. Origins and History of CSCW

Barley *et al* (2004) have pointed out that CSCW has been driven by accumulated empirical data and very little effort has been put into developing sound theories to underpin the field. However this is now changing and as more data is accumulated Barley *et al* have begun to build a situated theory of micro organisation. Furthermore the field of social psychology it is contended has been inadequately mined in the HCI and CSCW literature. This better understanding of the theory will lead to improved design for CSCW applications.

## 3.2. Basic characteristics of CSCW

CSCW takes HCI into the realm of collaborative working. This requires an analysis of collaboration in the workplace on top of HCI principles. The new features of collaboration and the way in which this is analysed and structured form the basis of this chapter.

Muller and Wu (2005) have remarked that within CSCW work is structured around five landmark entities which are:

- Documents including Drafts
- Dates and Calendars particularly start and end dates
- Events including the "kick off meeting" (first event)
- Roles and persons
- Systems and databases

These form the five landmark navigation points of any CSCW specification. However no clear conclusions are reached in this paper as this is a work in progress and the analysis is not yet complete.

Hawryszkiewycz, I. (1994) has looked at CSCW as a basis for interactive design semantics. He concludes that CSCW has a dual role; firstly as a research area in its own right and secondly as providing a language to describe the design process itself. Hawryszkiewycz, defines CSCW semantics and then describes how to use them to define a collaborative design process. Hawryszkiewycz, proposes five elements

- Artefacts (files, reports, documents, policies etc)
- Actors (a person in the organisation, each person can play many parts)
- Tasks (This is some well defined business function)
- Activities (the process for interactions between artefacts)
- Environments (provide the supportive structures for activities)

These elements are combined to model the design process using diagrams which are similar to Systems Analysis diagrams.

The analysis of these five elements would need to be carefully considered to see whether they are suitable to provide a model of CSCL later in the paper.

## 3.3. CSCW Theories and Principles

### 3.3.1. Design elements

Carroll *et al* (2006) has approached CSCW from a more primitive standpoint. He asks the fundamental question "What do Collaborators need to share in order to work together effectively?" the answers to this basic question involve the following kinds of knowledge and activity.

- Collaborators need to be assured that they partners are “there”
- They need to know what tools and resources counterparts can access
- -need to know what relevant information collaborators know
- -need to know their collaborators expectations attitudes and goals
- -need to know what criteria collaborators use to evaluate the outcome
- -need to know the moment to moment focus of their attention
- -need to know how the work is accomplished and evolves over time

Carroll’s analysis showed that long term collaborative work highlights four aspects of group work

- Establishment of common ground
- Performance in a community of practice
- Trust and social support through the formation of social capital (good will, willingness to communicate and work together, feelings of mutual endeavor etc.)
- Human development

Carroll consequently derives four design requirements each associated with one of the points above

- Public display of shared information
- Integration of data into community metaphors to facilitate analysis
- Aggregation of individual contributions into collective overviews to evoke trust and commitment
- Contrast of individual capabilities and roles to invite collaborators to perform beyond themselves

### 3.3.2. Design requirements

D’etienne (2006) has considered two characteristics of collaborative design taking place within cooperative work. The first concerns the way in which collaborators are interdependent according to the nature of the design problem and secondly the arrangement of designing cooperative work. Consequently D’etienne (2006) has suggested valuable new research directions in the following areas:

- The coupling of work and its organisation
- Informal communication and informal roles
- Awareness in distributed design
- Establishment of common grounds
- Perspective, clarification and convergence mechanisms in co-design

Following D’etienne this work will be looking in particular at the first and last but will also encompass the others.

### 3.3.3. Real Time Collaboration

Juby and De Roure (2002) have argued that real time collaboration requires more than just audio, video and data sharing, and have proposed two specific enhancements to provide a richness of interaction that is required for proper collaboration

Speaker identification. First hand experience of large meetings online has shown that a lack of perceptual clues (such as audio direction) has lead to difficulty in keeping track of who is actually speaking. In large scale meetings identifying the speaker is a difficult process and can involve scanning through dozens of video feeds to locate the correct person. The introduction of “directional audio” and “gaze direction” allows participants to track the protagonists. This is accomplished using sound source location techniques (SSL) where each speaker has their own microphone.

Participant tracking. Attendance lists are used to keep track of who is in the remote meeting rooms. This is required because not everyone is always on camera or displayed. Also people join and leave meetings mid-session, so it is important to have an up to date dynamic list of participants available.

It is felt that additional work needs to be done to identify precisely which enhancements are really important to provide a rich interaction environment for collaboration. The two mentioned by Juby and de Roure may or may not be the most important enhancements for every project, but at present there is no clear understanding of which are required for effective online collaboration.

### 3.3.4. Gestures

Karam and Schraefel (2005) have taken this a stage further by examining the role of gestures in HCI in order to see if this provides the necessary richness for effective collaboration. They provide a literature review of over 40 years of gesture based interactions which they then categorize into a taxonomy of gestures denoted by four key elements.

- Gesture styles. Five categories of gesture style are indicated which include gesticulation, manipulation, semaphore, deictic, and language gestures.
- Enabling technologies. This involves the ability to enable gestures on the computer and transmitted online and created through various input technologies such as a mouse movement, stylus, a touch surface, remote sensing, sensor gloves etc. All input devices can be used to provide some kind of gesture.
- Application domain. This concerns the area where collaboration takes place and online gestures may be employed. This could include virtual reality applications, CSCW interface interactions, Desktop style interactions with 3D worlds, immersive interactions (where the users' body is represented by an avatar), and robotic control environments.
- System response. This concerns the mode of output which is the end result of the gesture interaction and may include such elements as a visual output, an audio output, a CPU output.

Karam and Schraefel conclude that gestures are a natural, novel and improved mode of interaction. However it will take some time before they are incorporated as a standard feature into Microsoft Windows in the same way that say speech has been incorporated. They have demonstrated that there is a vast range of research in this area but very little application as yet. It still needs to be assessed how important gestures are for effective collaborative work, and it needs to be carefully considered whether we need to include gestures in the collaborative interface.

The problem is that if one tries to visualize instinctive and immediate gestures, the moment one thinks about using visualization the impact of immediacy is lost. Furthermore how can that be incorporated into an interface that can utilise intuitive expressions rather than deliberate statements, is not clear yet.

### 3.3.5. Unanticipated Use

Pekkola (2003) has considered the area of design for unanticipated use of artefacts. Under consideration is the design of a software application and concludes that support for common artefacts can be designed to a certain extent to make them more suitable for unanticipated uses.

This research is based on the idea that users did not always use applications as expected by their designers; instead they found alternative ways and reconstructed a use to match their work process. This has application to the design of user interfaces on software programs. They cite a particular case of an interface design which was set up to work in a particular way but its use was circumvented and improved by unanticipated shortcuts using the right mouse button instead. They conclude that "the search for common artefacts is a better starting point for analysis and design than a search for work sequence". This requires a revised process of design which involves "taking a step back, having an overview of the situation and making generalisations rather than concentrating on the sequence needed to perform a task.

## 3.4. CSCW Analysis Methodologies

### 3.4.1. Social Network Analysis

One method devised by Daniel B. Horn *et al* (2004) to analyse the impact of CSCW research involves the consideration of social network analysis. This is performed by viewing a network of authors as nodes and shared papers as links.

By this means it is possible to compare patterns of growth from one CSCW domain to another. See Figure 3 below:

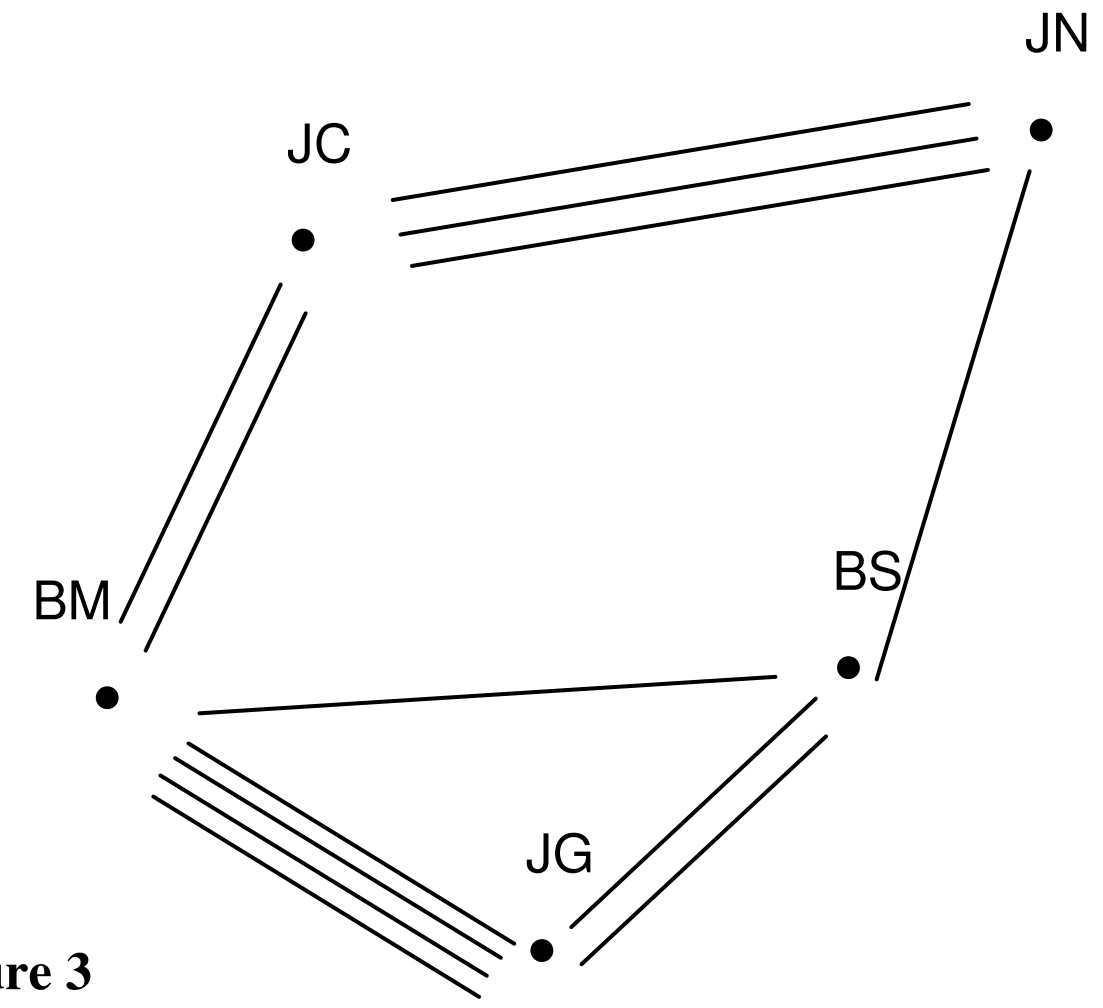

**Figure 3**

The initials refer to authors acting as nodes indicated by points and links represent the number of papers co-authored. It gives a graphical idea of the composition of the CSCW research community. This is called "social network analysis which is the primary lens to understand the patterns of collaboration".

"An individual with high centrality is potential influential because this person may link together many people who otherwise would not be connected". Horn *et al* (2004). Only recently with the advent of increased computing power has an analysis of very large communities numbering tens of thousands of members been possible enabling the depiction of ecologies of collaboration which might encompass an entire scientific discipline.

As a next step Horn *et al* have developed another view of the structure of interconnections and how they change over time. A dynamic network is constructed that represents the fields over a sequence of time intervals. By introducing the assumption that collaborative links persist for a certain time and then decay if not renewed a dynamic picture is generated. The time period chosen is five years and this link is removed if there is no further collaboration. This provides a clear picture of the succession of new generations of researchers and their movement into and out of a given field.

The raw data for this analysis has come from the database of HCI publications held by ACM at http://www.hcibib.org and is similar to the analysis performed by RVHH (2006b) who produced an analysis of the citation frequencies of HCI authors and their design principles. This method led to the construction of a taxonomy of design principles weighted according to the citation frequency of their proponents.

Horn *et al* have produced a number of results and suggestions for further study.

- Firstly they have shown that geographical proximity strongly affects collaboration and that successful collaboration seems to depend heavily on co-location since only four out of twenty have managed to maintain collaboration at a distance.
- In addition they conclude that co-authorship data is also useful to identify gaps in research which would benefit from collaboration.

Nurmela *et al* (1999) consider a social network analysis as a method to evaluate processes taking place in a CSCL environment. In particular they look at log files as the basis for data collection and analyse these to provide a perspective on how well the collaboration has worked. Social network analysis is focussed on uncovering the patterns involved in people interacting with each other. They conclude that log files can be used to evaluate the effectiveness of CSCL.

The concept of cohesion is introduced as a measure of the extent of direct interaction between individuals or actors. The aim is to find the most impressive actors in the field by using Freeman's degree, which measures the network activity of individuals. To give an idea of network analysis the following figure will illustrate the nature of the relationships.

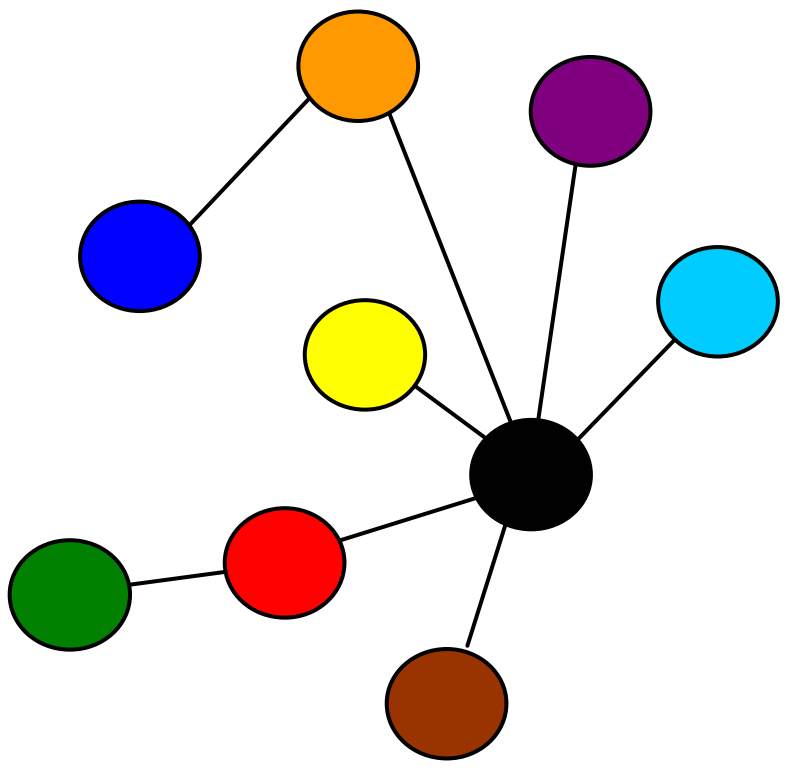

**Figure 4.1**

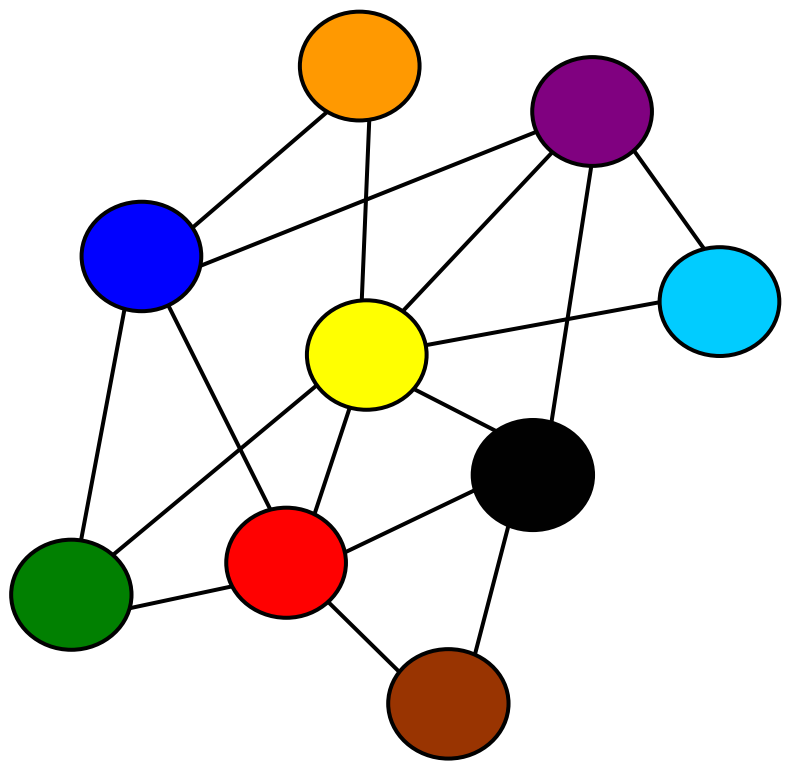

**Figure 4.2**

Each circle represents a collaborator and the black circle represents the tutor. Figure 4.1 shows that the tutor is the centre of almost all conversations and has all students around him. Students rarely relate to each other. Figure 4.2 on the other hand shows a distributed communication pattern where all collaborators communicate with a range of people and the tutor is just one member of the group. The second example shows a more balanced and healthy collaborative environment. Such a network analysis based upon the logs of collaborators may well be very useful to determine the successfulness of the collaborative learning environment.

## 3.5. CSCW systems and models

Dourish P. (1988) has developed a CSCW toolkit called Prospero that allows both synchronous and asynchronous activities which goes beyond the traditional toolkit range by being applicable to different situations and different demands.

Zann Gill (2002) has proposed the use of a "web tank", which he defines as a think tank on the web to support collaborative and innovative sharing of ideas. According to Gill these "web tanks"can serve as Petri dishes to culture the creative process" and they also allow invisible observers to study the environment from outside. This sharing of ideas on the web enables the development and testing of knowledge management strategies. They are in effect a new kind of intranet that is project based rather than institution based. In other words, use the web as a collaborative learning tool.

## 3.6. Review of CSCW Research Findings

### 3.6.1. Anonymity

Postmes *et al* (2001) have studied the social influences in computer mediated communication. Using a Social Identity model of Deindividualisation Effects (SIDE) they predicted that introducing anonymity to users would promote social influence within CSCW. This means that by allowing contributors to remain anonymous throughout their communications they are prepared to interact more and become more vocal participants and show a higher degree of influence within a group. They conducted tests with both anonymous groups and identifiable groups.

The implication for the Sim 20 laboratory is not necessarily clear cut. Social anonymity produces greater influence in larger groups where there would normally be reservation when contributions can be seen by a large number of people. However in paired work this anonymity may not have such a scale of importance. Postmes groups were of three or four people who were isolated from the other participants during the whole of the study who were all asked to do the same task. Postmes *et al* did not deal with groups of two.

Sassenberg and Postmes (2002) have taken the question of anonymity further. They performed two studies, the first of which showed that anonymity of group members enhanced the unity of the whole group while at the same time the interpersonal attraction of a group leader was diminished. The second study allowed participants to see pictures of other participants at random and not to see the real picture of each person. This preserved a kind of anonymity while testing to see if visual indication of other participants had any effect on the quality of communication. This second study showed that individual differentiation through showing digital photos (even though they were not representing the real students) did positively increase social influence and that individuality became more important again. The implications for the SIM20 laboratory are that maintaining a strict anonymity will increase social interaction and increased individual contribution whereas any form of individual differentiation (anything that distinguishes one individual from another) will tend to polarise contribution.

Spears *et al* (2002) concur with Postmes that isolation and anonymity in cyberspace produce more social interactions rather than fewer. People can be more outspoken online than they would be in real life which can lead to social repercussions if the anonymity is taken away.

### 3.6.2. Negotiation

Swaab *et al* (2004) have examined the issue of how technology interacts with the role of collaborative negotiation. They produced two findings

Negotiation support systems should stimulate a common cultural identity among the individual participants

Negotiation support systems should provide information to develop shared cognition among negotiators

### 3.6.3. Social Identity

Watts and Reeves (2005) have pointed out that email lacks social sensitivity and can be detrimental to communication by fostering misunderstanding. They propose to develop a more effective means of communication by integrating common ground, social identity and privacy, but this has not yet been accomplished.

# 4. Computer Supported Collaborative Learning (CSCL)

## 4.1. Origins and History of CSCL

CSCL has grown out of CSCW. According to the University of Texas CSCW is defined as “a computer based network system that supports group work in a common task and provides a shared interface for groups to work with”

The following table indicates the main differences between CSCW and CSCL.

| CSCW | CSCL |
|---|---|
| Focuses on communication techniques | Focuses on what is being communicated |
| Used mainly in a business setting | Used mainly in an educational setting |
| Purpose is to facilitate group communication and productivity | Purpose is to support students in learning together |

**Table 1**

## 4.2. The Basic Characteristics of CSCL

In his review paper Bannon (1989) indicates that there are two entirely different perspectives on Computer Supported Collaborative Learning (CSCL). The first view is that CSCL is an umbrella term that simply means whatever people who work in the field say it means. In this sense it is useful only as a generalised terminology. The second view is that CSCL is a specific problem area for researchers with four component parts:

Learning- This is seen as an activity that takes place in a wider context than a classroom and involves the everyday social practices of people during which learning occurs and the situation it springs from (Lave and Wenger, 1990)

Collaborative learning – The role of others in the learning process has been highlighted by Vygotsky (1978) and his key concept of the zone of proximal development (ZPD) as the area of overlap between the inexperienced and the experienced in which learning occurs. Issues that need to be considered will include

- The nature of the collaborative task
- Who are the collaborators (Peers, Teacher- Student, Student – Computer)
- Number of collaborators
- The previous relationship between collaborators (how much shared history and vocabulary)
- The motivation for the collaboration (work requirement, money, mutual interests)
- The setting of the collaboration (Home, Workplace)
- Conditions of the collaboration (Co-present, separate)
- Time period of the collaboration (Minutes, Hours, Days)
- The chronology of the collaboration (synchronous or asynchronous)
- Supported collaborative learning; an analysis of what tools are required to provide the environment and the mechanisms for collaboration.

Computer supported collaborative learning. The computer brings a new dimension to the process of learning and can introduce a number of new features: It can-

- allow the simulation of situations impossible in the real world
- record student actions to improve problem solving
- reify (make manifest) the process of thinking
- help create functional learning environments
- be used as a simple data gathering tool
- be seen as a tutor with whom the pupil interacts and even collaborates
- provide assistance in the coordination of joint activities

In short CSCL facilitates the learning process through a number of applications including email, computer conferencing, bulletin boards, local area networks, and hypermedia. It is Bannon's (1989) contention that the best way to regard computers in the CSCL process is as an enabling medium through which partners can organise and accomplish activities. The computer provides a space to work in which others can organise their activities.

Lipponen (2002) defines CSCL as being focussed on "How collaborative learning supported by technology can enhance peer interaction and work in groups and how collaboration and technology facilitate sharing and distribution of knowledge and expertise among community members". He goes on to define collaboration to "refer to any activities that a pair of individuals or a group of people perform together". This definition is improved by bringing in the twin ideas of "co-construction of knowledge" and "mutual engagement of participants".

Alternatively another definition which has come from educational research sees collaboration as a human activity essential for cultural development where participants work together but also includes the idea of achieving a shared goal.

The nature of collaboration according to Dillenbourg (1999) concerns four aspects of learning

- A situation (a gathering of people of similar status)
- The interactions (between group members)
- Learning mechanisms (tools for collaborative sharing)
- The effects of the learning process (the need to measure collaborative learning)

Dillenbourg (1999) further characterizes a situation (see Figure 5) as collaborative if it meets the following four criteria. Peers must

- be more or less at the same level
- be able to perform the same actions
- have a common goal
- work together

A degree of symmetry within the role of collaboration is identified.

- **Symmetry of action** (the extent to which each collaborator has the same range of actions)
- **Symmetry of knowledge** (the extent to which collaborators possess the same level of skills)
- **Symmetry of status** (the extent to which collaborators have the same status with respect to their community)

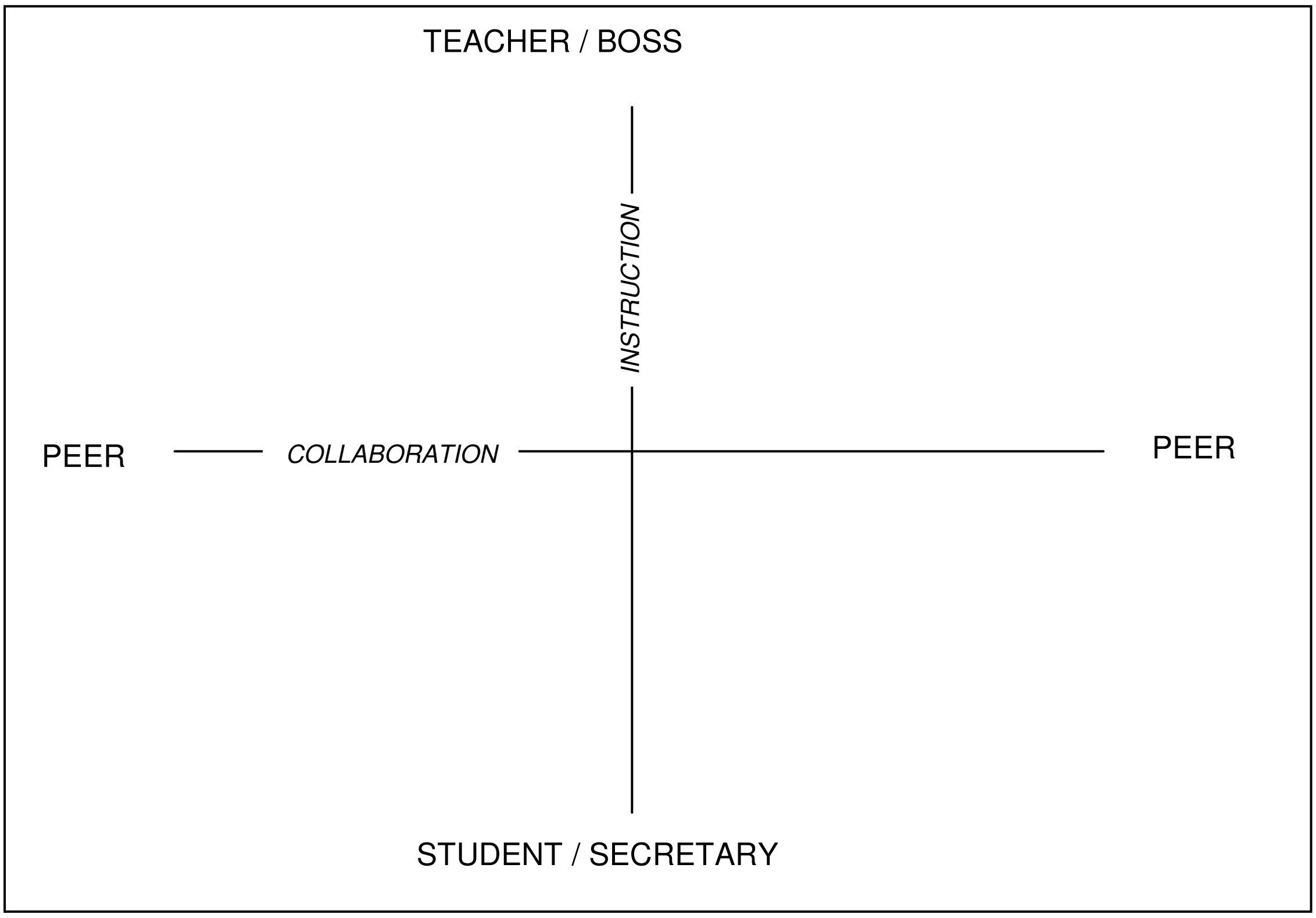

Figure 5

These symmetries are dependent upon the individual and differ from one collaborator to another and they may also vary over time. For the purposes of learning CSCL a slight knowledge asymmetry is useful for encouraging the peer learning process.

Dillenbourg (1999) goes on to specify the defining criteria for collaborative interaction. These involve three areas:

**Collaborative situations should be interactive.** The degree of interactivity is not defined by their frequency but by the extent to which they influence their peers thinking and reasoning. This is not a simple matter to measure and Dillenbourg defines it as a major challenge for collaborative learning research to produce operational criteria which will define the amount of cognitive interaction between participants.

**Collaborative situations should be synchronous**. This is the equivalent of “doing something together” where the speaker expects the listener will wait for his message and will process the message as soon as it is delivered. However this may be interrupted by the medium (on line chat, Skype, messaging etc) if it is too slow or does not work appropriately. The medium used needs to ensure that the subjective feeling of synchronicity of reasoning is maintained.

**Collaborative situations need to be negotiable**, one partner must not impose his view entirely upon the other, but must to some extent argue the case, justify, negotiate and convince. The implication here is that collaborative dialogue will be more complex than other types of dialogue such as lecturing. Negotiation can sometimes founder on misunderstanding rather than on disagreement. Therefore Dillenbourg sees it as a central concern for research in collaborative learning that the elimination of misunderstanding is implemented in the interface. This requires the elimination of ambiguity if possible. Ambiguity free (or noise free) communication protocols in multi agent systems are a prime concern of collaborative learning research. Example: When two partners misunderstand each other they need to construct explanations of their point of view as well as justification and reformulation of their statements. All of these activities can lead to learning.

### 4.2.1. Individual learning

Dillenbourg moves on to consider the differences between individual learning, paired learning and collaborative learning and identifies the following four elements as being central to individual learning:

- Induction – This is the process whereby a representation of the world is build up by the individual on the basis of what they have discovered so far. It is an extrapolation from particular cases to a general rule.
- Cognitive Load - The degree of work taken on by an individual can reach a high saturation point to the extend that individuals may reach a limit faster than if they were working collaboratively.
- Self explanation – Individual cognition is based on the concept of explanation based learning where each individual has to rationalise a meaning of what they learn to themselves in order for it to become meaningful to them.
- Conflict – This is part of Piagetian theory in which central to the development of an individual is the discovery of conflict within the social environment and the resolution of that conflict leads to learning. Example: The knowledge that stars are large objects conflicts with the view of the stars in the sky. Resolution of this conflict leads to learning about distance.

### 4.2.2. Paired and collaborative learning

Each of these individual learning elements are also applied to paired learning and collaborative learning

- Induction – It is well known that pairs draw more complex representations because joined drawing has to integrate the common representations of each individual Dillenbourg (1999). This process is inductive since each learner must induce patterns from the expressions of his partner
- Cognitive load – In collaboration the division of labour reduces the amount of processing by each individual. This explains why group members can improve their skills in regulating their partner's processes and influencing their actions.
- Self explanation – This can be re- exported into a collaborative setting by each partner providing an explanation to the other which produces learning on one side and the consolidation of learning on the other.
- Conflict – This concept has been exported to the case of social interactions where a discrepancy between the viewpoints of two collaborators may lead to conflicting positions and where the resolution of that conflict can lead to a greater understanding for both parties.

Berger *et al* (2001) have designed a CSCL environment for use across countries with different social, economic and cultural backgrounds. They based their design on seven principles:

- **Exploit differences between participants**. Differences in viewpoints between co- learners provide the dynamics for social interaction. Participants may have natural differences (social, cultural, economic etc.) or artificial differences where CSCL environments deliberately provide participants with divergent information in order to create interesting dynamics. For instance different groups might be provided with different case studies for the same issue. Bridging these differences is designed to lead to deeper understanding.
- **Structure collaboration through a scenario**. Collaborative learning does not automatically guarantee any improvement in cognitive learning. This depends on the extent to which participants engage in rich interaction. The specific structure that creates rich interactions is a scenario which is based on roles and phases.
- **Regulate the collaboration**. Collaborative processes by themselves do not guarantee effective learning unless carefully regulated. The scenario must therefore also define the role of the teacher who does not give solutions but facilitates the discussion only. This may take place through face to face meetings, electronic forums or live conferences.

- **Reify the scenario.** To facilitate navigation they designed an environment around a graphical representation of the scenario. This representation is an on screen time line indicating stages in the tasks to be performed with a triangular cursor indicating the current phase.
- **Use a hybrid scenario**. Use is made of both face to face teaching as well as web based environment which forms a hybrid integrating both kinds of activity.
- **Build around a knowledge space**. Plain HTML is insufficient. CSCL environments require a knowledge space which defines roles and enables the students to interact and co- construct.
- **Negotiate knowledge space.** This involves a set method for coming to a consensus between participants and involves three phases.
    - **Phase 1** involves a sending out of a questionnaire and asking the participants to list their priority issues.
    - **Phase 2** takes these results and formulates the top ten items. These are sent out into a second questionnaire and participants are asked to rank them in order of importance.
    - **Phase 3** consists of asking the participants to comment on the final order.

Berger *et al* have concluded that these are the essential design elements necessary to provide innovative collaboration that exploits the differences between the participants while building on their strengths.

Bouras *et al* (2002) takes the situation a stage further and has set up a virtual community for e learning. According to Bouras *et al* (following Dillenbourg P. (1999)) an e learning community is characterised by the following features:

- The environment should be explicitly designed for the task and have common features with a physical space and support different learning scenarios.
- Educational interaction should change virtual space into a communication space.
- Learners should not be passive and should be able to interact.
- The system should be able to incorporate other technologies e.g. spreadsheet, database etc.
- E learning environments should be based on templates and be able to handle users' materials and interactions.
- The environment should offer both synchronous and asynchronous communication including chat, audio, email, forums, shared objects, application sharing, and gestures.
- The environment should be populated by users who are represented by 3D avatars.
- The environment should be aesthetic and easy to use

Bouras *et al* also identify the main entities taking part in the collaboration as the learners, the tutors, and the moderators together with shared objects and educational material. The learners have a limited user interface with reduced capability but allowing full interaction. The tutor interacts with the whole system and can perform any task. The moderator participates only in the collaborative activity in real time and determines roles activities and participative permissions. Bouras *et al* also identify what they consider the essential learning tools which include

- A 3D whiteboard
- A 2D whiteboard (i.e. Microsoft's net meeting whiteboard)
- A 3D library which contains links to WebPages, with educational material (in Bouras' case this took the form of a library where each book on the shelf represented an appropriate link and which supported the dynamic addition and removal of books).

Bouras also required as a necessary condition of operation the ability to:

- Lock and unlock objects which prevents more than one user accessing a certain object at a time

- Ability to expel a participant which becomes necessary when a learner becomes annoying or seeks to prevent the smooth progress of a lecture. The interface contained an **expel button** for this purpose.

Bourguin, G. and Derycke, A. (2001) have pointed out that the integration of computer supported collaborative learning CSCL into the wider framework of a virtual learning environment is a difficult challenge. This is because CSCL tools are mostly designed for stand alone use and interoperability within a VLE is difficult to accomplish if not impossible. The implications of this are that a dedicated learning environment for collaborative purposes should be constructed from the bottom up.

According to the University of Texas  CSCL systems are categorised according to the time location matrix as illustrated below:

| | Same Time | Different Times |
|---|---|---|
| Same location | Back Channel communication<br>E-learning laboratories | Bulletin Boards<br>Forums |
| Different location | Video Conferencing<br>Chat<br>Whiteboards<br>Messaging<br>E learning laboratories | Email<br>Forums<br>Web logs and Journals |

**Table 2**

## 4.3. CSCL Theories and Frameworks

John Carroll (1990), in his book the Nurnberg funnel has argued that “the learner, not the system, determines the model and methods of instruction.”  By this he means that the most rapid achievement of learning does not come from drill and practice techniques nor from standard training methods but rather from instruction via error recognition and recovery, and the study of peoples learning problems and how they are solved. The Nurnberg Funnel refers to a legendary device that made people wise very quickly. By designing minimalist instruction methods based on what learners do spontaneously; a much faster method of learning that outperforms the standard approach in every relevant way can be achieved.

Lipponen claims that there are two mechanisms which promote learning in CSCL. The first comes from Piaget who said that children who socially interact with others have “conflicts of thought”. This “shock of our thought coming into contact with others” results in the construction of new conceptual structures. (They learn from the differences of each other’s ideas)

The second mechanism is based on Vygotsky’s ideas that people who engage in collaborative activities can master something which they could not earlier do as individuals. People gain knowledge as a result of reflecting on their collaborative activities. According to Vygotsky therefore learning is more to do with participation in the process of collaborative construction than it is to do with an individual learning from his differences with other people. (Learning from the collaborative experience).

### 4.3.1. Fundamental Educational Theories 1: Piaget – the individual

Kirschner and Gerjets (2006) have highlighted the importance of the individual in the learning process. “This new age of mass individualisation has led to demand- led learning”. Each learner is matched to their own characteristics, learning styles, knowledge and goals and is linked with specific learning materials and approaches provided by online learning environments. The adaptive provision of learning materials is seen as the best approach to helping individual needs where software agents use and select appropriate materials which are optimally suited to the student's performance on previous tasks. This means that every learner can have their own individual learning plan, and be taught in a way that suits them specifically. The result of this approach according to Kirschner and Gerjets is that learning becomes not only more effective but also more enjoyable.

### 4.3.2. Fundamental Educational Theories 2: Vygotsky – the social

CSCL is based on the assumption that individuals are purposely seeking and constructing knowledge within their social environment and that a computer system can facilitate that learning. The role of the computer is

- To act as a cognitive tool that teams individuals with technology
- Shares the labour of communication during group processes
- Functions as a scaffolder to provide resources
- Assists in modifying individual's cognitive abilities
- Offloads part of the cognitive process to enable the individual to concentrate on learning

Much of this is based upon Vygotsky's sociocultural theory of learning which teaches that individuals learn first through interaction with others in a social environment rather than working things out for themselves.

Vygotsky's ideas are that an experienced person can help an inexperienced person only if there is an overlap between their knowledge areas. This equalled the Zone of Proximal Development ZPD where learning from others takes place and which is supported by CSCL.

In addition the constructivist approach to learning suggests that knowledge is gained by an individual constructing it through their own experience of the real world. Bruner (1960) has argued that this model of learning emphasises knowledge construction through the active participation in social and cultural contexts which is supported by CSCL.

Daniels (2001) argues that "unless we understand the ways in which possibilities for learning are enacted within institutions we will be frustrated in our attempts to really raise standards" this also follows the work of L.S. Vygotsky who considered the capacity to teach and benefit from instruction is a fundamental attribute of human beings. The effect of this on our implementation of an analysis of a collaborative learning environment gives equal weight to both the learning and teaching that takes place there.

### 4.3.3. Problem based learning

A third component to CSCL is problem based learning which involves

- Developing a scientific understanding through real world cases
- Developing reasoning strategies
- Developing self directed learning strategies

The importance of collaborative discourse in CSCL is that it activates prior knowledge which assists the acquisition of new knowledge through observation and guided practice.

Kligyte, (2001) in work submitted for his MA thesis discusses the issues of the computer interface as a common boundary shared by a group of people. His work is based on the "progressive enquiry pedagogical model" which adopts the view that new knowledge is not assimilated but constructed and transformed through problem solving. This is very similar to the constructivism of Bruner. As a consequence CSCL needs to take into account this ability to construct knowledge whereas CSCW is, in his view, mostly focussed on the simple problems of efficient document management.

In summary, according to these theories the main role of the teacher is not to impart knowledge but to equip the students with strategies to become independent thinkers and lifelong learners. Any CSCL environment needs to develop systems which allow for the development and support of metacognition (A person's knowledge of what they know and how they learn) and problem solving skills development.

#### 4.3.4. CSCL/ HCI Frameworks

Stahl (2002) examines the theoretical framework for CSCL and presents four hypotheses or claims that:

We should use the term “knowledge building” rather than “learning” in the field of collaboration. The activities by which shared knowledge is constructed is more easily analysed in terms of group perspectives.

Collaborative knowledge building comes from the intertwining of the group perspective and the personal perspective. It is important not to give too much or too little attention to the role of the individual mind in group learning.

Construction of knowledge occurs through the mediation of artefacts. These can be as simple as words or gestures and as complicated as digital media or any kind of computer file. It is the sharing of these new artefacts that communicate new knowledge.

The process of collaborative knowledge building can be captured by e.g. video and rigorously analysed to make visible the knowledge building activities.

In summary, according to Dillenbourg (1999) a theory of collaborative learning must concern these four items, Situation, Interaction, Processes and Effects. These can be highlighted in the following table:

| Four Aspects of Learning | Collaborative Learning |
|---|---|
| **1 Situation** | **Peers must be**<br>**- at the same level**<br>**- must be able to perform the same actions**<br>**- must have a common goal**<br>**- must work together** |
| **2 Interactions between group members** | **Symmetry of action**<br>**Symmetry of knowledge**<br>**Symmetry of status** |
| **3 Processes** | **Interactive**<br>**Synchronous**<br>**Negotiable** |
| **4 Effects** | **Measurement of groups not individuals** |

**Table 3**

## 4.4. CSCL Analysis Methodologies - Measuring the Effects

Michael Hannafin (2006) raises the question of the proper methodology in the study of computer supported learning CSCL. Hannafin points out that there is a debate concerning the proper or correct methodology to use and there is much disagreement in the field. He quotes a number of sources which compare present research in CSCL to pseudo science while he quotes others who applaud methodologies as scientifically valid. He concludes that there are wide differences of opinion as to the proper study of technology and learning. By applying Pasteur’s quadrant to differentiate research endeavours he draws the conclusion that three quadrants are relevant to research in CSCL

| COMPUTER SUPPORTED LEARNING (According To Hannafin) | | |
|---|---|---|
| | **Theoretical** | **Applied** |
| **Fundamental Understanding** | Foundation Research<br>Cognitive Science<br>Perception<br>Systems Theory | Theory-building Research<br>Learning Sciences<br>Pedagogy |
| **Complex Understanding** | | Applied Research<br>Instructional Design<br>Instructional Technology<br>HCI |

**Table 4**

According to Hannafin the three areas of research involve first the fundamental theoretical areas including cognitive science, perception and the functioning of the human mind. Secondly the fundamental applied areas which involve educational theory and pedagogy. Finally in the complex applied area there is the research on how the instructional technology and HCI functions. In our opinion, however, it does not seem that this is a helpful analysis of the CSCL research areas as it overlooks areas of collaboration and introduces a pseudo difference between fundamental and complex understanding.

A suggested and more helpful alternative analysis of CSCL research areas is based upon a distinction between the separate approaches of educational, HCI and collaborative theories

| **COMPUTER SUPPORTED LEARNING (According TO RVHH)** | | |
|---|---|---|
| | **Theoretical** | **Applied** |
| **Educational** | Foundation Human Research<br>Cognitive Science<br>Perception<br>Systems Theory | Educational Theory Research<br>Learning Sciences<br>Pedagogy<br>Constructivism<br>Situational |
| **HCI** | Foundation<br>Computer Research<br>Usability<br>HCI Theory | Applied HCI Research<br>Instructional Design<br>Instructional Technology<br>Ergonomics |
| **Collaborative** | Foundation<br>Collaborative<br>Research<br>Interaction Theory<br>Social Behaviourism | Applied Collaborative<br>Research<br>SIM 20<br>VLE<br>ALICE |

**Table 5**

### 4.4.1. Collaborative Toolkits

Ørngreen *et al* (2004) in their working paper have analysed the important features of a pedagogical learning environment and have concluded that the necessary collaborative tools must include the following:

- Discussion forums
- Collaborative features (the ability for subgroups to create their own private conference areas)
- Chat tool
- Audio visual support 1 (dissemination of pre- produced material and broadcasts)
- Audio visual support 2 (live transmissions)
- Whiteboard (to demonstrate experiment or software utility)
- Different access rights for different users

Bachler *et al* (2004) have considered tools necessary for collaboration in the semantic grid. They have identified four significant tools

- Instant messaging and presence notification ( buddy space)
- Graphical meeting and group memory capture (compendium)
- Intelligent to do lists ( process panels)

- Meeting capture and replay

These four tools when integrated into a collaborative environment and using a shared ontology are sufficient to promote process tracking and navigational resources before, after and during a meeting.

Within the domain of the semantic web a vocabulary is necessary to express relationships between entities. These entities will include

- Individuals
- Projects
- Locations
- Documents and Publications

The reference ontology needs to be suitable for encapsulating

- The meeting event itself
- The attendees to the meeting
- The projects which will be the subject matter of the meeting
- Documents associated with the meeting including multimedia

Each event is broken down into sub events within the meeting structure such as; giving a talk, sending an email, publishing a paper etc.

Kligyte (2001) has laid out a scheme for the toolkit required for using CSCL. He splits these functionalities into o two areas asynchronous and synchronous communication.

- Asynchronous communication
- Document sharing spaces (a file management functionality)
- Discussion spaces (bulletin boards with threads)
- Argumentation tools (facilities for posted notes)
- Whiteboards (sharing graphical information)
- Synchronous communication
- Whiteboards and presenters
- Text based chat
- Moos and Muds (Multi user environments)
- Avatars (2D or 3D representations of participants)
- Audio chat and video conferencing

Leinonen *et al* ((2002) have developed a collaborative discovery tool (CoDi tool) that enhances knowledge building in their Fle2 learning system. The Codi tool is an additional facility which helps to collect and manage knowledge and enquiry. It also provides for the marking of key ideas and the data which is produced from this can be used to create visual representations of the database which help participants to build on recognised ideas. The questions this raises concern the marking of students work and how that should be organised. Further work needs to be done on how marking facilitates knowledge advancement and whether the results should be shared not only by tutors but also by students. Leinonen *et al* do not yet have an answer to these questions.

#### 4.4.2. Content Analysis

Weinberger and Fisher (2006) have been looking at a framework to analyse knowledge construction in CSCL. This is normally based on an analysis of the written transcript of the learner. However, more recent approaches facilitate knowledge construction with collaboration scripts. These scripts or scenarios set out for the student a particular course to follow which is designed to lead to the process of knowledge construction. Weinberger and Fisher propose a new multi dimensional approach by analysing the influence of scripts on the dialogue between students in terms of a four dimensional process as follows:

- **Participation** – the quality of participation concerns both the quantity of contribution to the discourse and also whether students participate on an equal basis
- **Epistemic** (how do we acquire knowledge) – This analyses the content of the learners' contribution and refers to how learners work on constructing knowledge. For this purpose a distinction is made between the amount of "on task talking" and "off task talking". The former has been shown to positively increase knowledge acquisition.
- **Argumentative** – This requires the analysis of student dialogue for the construction of arguments and the construction of sequences of arguments. The number of argumentative statements must be measured again the measure of non-argumentative statements.
- **Social mode** – This describes to what extent learners refer back to the contribution of their fellow learners. This is found to relate to knowledge acquisition. A distinction is made between
    - externalisation  (learners make contributions without referring to anyone else's contribution) ,
    - elicitation ( learners aim to ask questions to receive information from others),
    - quick consensus building (learners accepting the views of others not because they are convinced but simply to move on)
    - integration oriented consensus building (learners operate on the basis of the reasoning of other learners by actively changing their own views
    - conflict oriented consensus building (learners face criticism from others and modify their views to form a better argument or take a better position). This is one of the most important components in collaborative learning.

The process is one where the transcripts of the students are collected; they are then segmented on a micro and macro level. Macro level segments represent how learners connect principles and concepts whereas micro level segments represent a learner grasping a single concept. Once the transcripts have been segmented the segments are coded with a set of categories according to the four dimensions above.

Weinberger and Fisher draw out a number of limitations of their system

- They do not know how useful the framework is outside of CSCL
- The application of the framework is a challenge due to the enormous workload of analysing transcripts on multi dimensions on a micro and macro level, and suggest that the use of a computer tag generator might streamline the process.

De Wever *et al* (2006) have also focussed upon the analysis of transcripts of asynchronous computer mediated discussion groups in educational settings. They begin by pointing out that although content analysis is frequently used there are no established standards. The differences of approach stem from different theoretical bases, the amount of information about validity and reliability and the choice of the units of analysis. They present an overview of different methodologies and discuss fifteen instruments. Their analysis focuses on text based CSCL tools.

| AUTHOR | THEORY | DIMENSIONS |
|---|---|---|
| Henri (1992) | Cognitive & Meta cognitive | 1 Participative<br>2 Social<br>3 Interactive<br>4 Cognitive<br>5 Meta Cognitive |
| Newman et al (1995) | Critical thinking | 1 Relevance<br>2 Importance<br>3 Novelty<br>4 Outside knowledge<br>5 Ambiguities<br>6 Linking ideas<br>7Justification<br>8 Critical assessment<br>9 Practical utility<br>10 Width of understanding |
| Zhu (1996) | Vygotsky-Cognitive and constructive learning | 1 Social interaction<br>2 Answers<br>3 Information sharing<br>4 Discussion<br>5 Comment<br>6 Reflection<br>7 Scaffolding |
| Gunawardena (1997) | Social Constructivism | Phase 1<br>1 Sharing and Comparing<br>2 Observations<br>3 Opinions<br>4 Agreement<br>5 Examples<br>6 Clarifications<br>7 Problems<br>Phase 2<br>1 Discovery of disagreement<br>2 Exploration<br>Phase 3<br>1 Negotiation<br>2 Agreement<br>3 Proposals<br>Phase 4<br>1 Testing<br>2 Modification<br>3 Synthesis<br>4 Co-construction<br>Phase 5<br>1 Agreement |
| Bullen (1997) | Critical thinking | 1 Clarification<br>2 Assessing evidence<br>3 Making and judging inferences<br>4 Strategies and tactics |
| Fahy et al (2000) | Social network theory | 1 Number of participants<br>2 Density of links to total<br>3 Intensity of links to each other<br>4 Flow and direction of information |
| Veerman (2001) | Constructivist | 1 New ideas<br>2 Explanations<br>3 Evaluation and Criticism |
| Rourke et al (1999) | Community of Enquiry | 1 Cognitive Presence (learners)<br>2 Teaching Presence (teachers) |
| Garrison et al ((2001) | Community of Enquiry | 1 Initiation<br>2 Exploration<br>3 Integration<br>4 Resolution |
| Anderson et al (2001) | Community of Enquiry | 1 Designer<br>2 Facilitator<br>3 Subject matter expert |
| Jaervaela (2002) | Social Constructivism | 1 Theory<br>2 New Pont/Question<br>3 Experience<br>4 Suggestion<br>5 Comments |
| Veldhuis (2002) | Social Constructivism | 1 Participation and Interaction<br>2 Learning activities<br>3 Constructed knowledge |
| Lockhorst (2003) | Social Constructivism | 1 Participation |
| | | 2 Content<br>3 Links between ideas<br>4 Information processing<br>5 Procedural information |
| Pena-Shaff (2004) | Social Constructivism | 1 Participation<br>2 Interaction rates |
| Weinberger and Fisher (2005) | Social Constructivism | 1 Participation<br>2 Epistemic<br>3 Argumentative<br>4 Social mode |

**Table 6**

De Wever *et al* conclude that empirically validated content analysis instruments are still lacking. De Wever *et al* call for replication studies that focus on the validation of existing instruments.

#### 4.4.3. Scripts and Scenarios

Dillenbourg (2006) has discussed the role of CSCL scripts and their role in measuring the performance of CSCL operations. Scripts (or scenarios) are a method of breaking down a classroom activity into distinct task, phases, roles and specific types of interactions. “Scripts are a modern version of instructional engineering applied to social settings”

In a number of ways there are similarities between scripts and lesson plans where each may include activities (e.g. lecture, de-briefing, collaboration etc.) and set phases (e.g. reading papers, writing a synthesis). However, they differ in that social interactions such as argumentation, explanation, justification etc. would be the key learning mechanisms in script based activity, whereas tutor input such as lectures and seminars are the key learning mechanisms in lesson plans.

Miao *et al* (2005) propose a particular CSCL scripting language and develops the conceptual framework. In addition two developed CSCL script authoring tools are presented. For Miao *et al* a CSCL script consists of a set of roles, activities, transitions, artefacts and environments and is formalised to the extend of defining a clear syntax and language for creating scripts. These will have attributes such as learning context, learning objectives, pre requisites, design rationale, duration, target audience etc. and enables a clear codification of a particular learning scenario. In the absence of any experiments they cannot offer a conclusion as yet.

Laister and Koubek (2002) advocate the social context of learning as needing to be the centre of attention and not merely an add-on feature of the e learning process. They further suggest that innovative learning platforms need to be built on collaborative learning scenarios. Scenarios require a paradigm change as far as current learning approaches are concerned.

James Dalziel (2003) has championed a particular process called “learning design” which captures the process of education rather than its content. This is a modular approach which describes sequences of collaborative learning activities that can be reused in different contexts and incorporated into a learning activity management system. Once a sequence of activities has been designed to teach one subject area the same activities might then be employed to teach another area. This methodology is strongly akin to the creation of learning scenarios. It needs to be seen whether this is a suitable approach for the scenario creation of simulators.

## 4.5. CSCL Systems and Models: Particular e Learning Environments and collaborative learning e-labs

A number of groups have set up learning e-labs and some pertinent ones will be reviewed in this section.

Bardeen *et al* (2005) from Fermilab have set up grid computing technologies to support the collaborative learning of students investigating cosmic rays. Their students use web browsers and a custom interface which enables them to perform the following tasks:

- Gather and upload science data to the e-lab portal
- Analyse those data using standard techniques

- Virtual data transformation
- Managing workflow
- Metadata cataloguing indexing,
- Data product provenance and persistence
- Job planning
- Sharing results
- Question other collaborators asynchronously
- Access other students' work
- Model large scale scientific collaborations
- Provides additional tools for teachers to guide student work

Fermilab are at present designing an assessment procedure to determine the efficacy of these tools in supporting the learning process. This takes place in two phases

- Phase 1 – Improving and testing the web interface tools. This involves iterative testing which watches teachers and students while they use the site
- Phase 2 – Involves an assessment of the learning that takes place. Activities include building online tests both before and after activities have taken place. Also student observations, think-aloud interviews, review of student work, examination of e logbooks and a discourse analysis on student use of communication and collaboration tools.

Harper *et al* (2004) have created interfaces which are contextually aware and adaptive to the way that humans naturally communicate and interact. These new "dynamic meeting environments" are suitable for collaborative research and contain perceptive assistive agents (essentially human ergonomes) which function as "friends and advisers" to the user in the real world collaborative setting. This setting is a 3D virtual representation of a seminar room and provides a rich environment for group activities with a range of features that will allow presentations, briefings etc. The key to real and full collaboration in an on-line environment is according to Harper *et al* to make the on-line environment as an exact a copy of the real environment as possible. This will give students the feel and full range of activities available in a real environment. Whether it is necessary to reproduce exactly the full real time learning environment in its entirety or whether a subset of those elements would be sufficient to maintain full collaboration needs further investigation.

Miettinen *et al* (2005) have introduced a system called OurWeb and they use it to demonstrate intelligent tutoring in a structured setting. This is claimed to be an intelligent e learning system which is designed to provide highly structured lessons under mostly automated control. This takes the form of adaptive sequencing which personalises the course materials to the students' individual need. OurWeb provides a framework of joint activities and the students are required to rely on each other for sources of information and instruction. OurWeb encourages them to explain the subject to each other. It does this by acting as a proxy between the users' browser and the web which provides additional content and a pop up menu for interaction and communication. There is also a shared document pool and documents are produced incrementally with students gathering feedback from others along the way. There is also an integrated Wiki which enables groups to collaboratively author web pages. Other features include annotation to web pages allowing artefact centred discourse, personalised desktops, an interface for sending email, and the ability to make comments on other students' comments.

Hosoya *et al* (1997) developed a collaborative educational system in which students at distant locations share a virtual 3D space. Students can move around within the space observing different objects from different viewpoints. In this environment users can respond to the actions of their collaborators and manipulate objects cooperatively.

The environment consists of five features

- A virtual shared space identified as a classroom (students can only exist in one room at a time but can move from room to room.
- 3D objects and educational materials. These can be shared by all students in the room.

- Avatars which represent the collaborators and allow students to recognise each other and to see in which direction they are looking and where they are situated at any given point.
- Real time conversation where collaborators can use either text chatting or voice- video conferencing.
- Cooperative operation of the virtual objects. This enables them to rotate objects, transform the shape, colour etc.

Whereas many shared learning environments may consist of the first four very few systems provide the last feature. Evaluation of this prototype system suggests to the author that it has a promising performance level but no details of this are given in the paper

Sendova *et al* (2004) have set up a virtual laboratory Toon Talk for collaborative e learning for young learners. The programs in Toon Talk take the form of animated robots which can be named, picked up and trained to form a sequence of steps. The Toon Talk weblab includes three elements

- Information exchange between computers
- A web report for participants to describe and share their ideas
- An environment to mediate the collaborative learning activities over distance.

This has proved to be valuable in teaching young people to collaborate with others and to be self critical of their work.

Silva and Liesenberg (2000) have studied a synchronous user interface where all the objects being shared can be viewed from many different locations and where users interact with each other in real time. They consider four different kinds of collaborative relationship

- Stand alone – does not interact with any other partner in the system
- Mediator – Partners interact only through the mediator
- EKEO – Everyone knows each other and relates together
- MEKEO –Mediator and everyone know each other where partners can request services from the mediator or from each other

They set out a methodology for designing a synchronous user interface that has five steps

- HCI design – This must be carried out in its own domain with independent HCI experts determining the system development process
- Support for user participation – This takes the form of collaborative design modelling where the user is an active member of the development team and describes their views in terms of their tasks
- Support for abstraction – This is the process whereby we identify only the important aspects of the design
- Support for re-usability – This allows a product to build a new version with the minimum expenditure by utilising earlier models
- Tool support – tools can be used to explore a notation and to directly develop prototypes of the design

Walters *et al* (2006) have considered the challenges involved in teaching the subject of distributed computing. They have set up an M-Grid framework which mimics the core features of a distributed computer system. This involves three parts

- A computing system that accepts tasks from outside users
- A distribution system that sends the tasks to available machines for processing
- A collection system to pass the results back to the users

In practice the grid network is a collection of computers that can be used in parallel to process computing tasks. The danger of accepting executable software from other computers on the system is a problem that requires a high degree of security to be in place. However, the M-Grid system makes use of Java applets running in a web browser which implements a sandbox constraining the action of the applets and preventing damage to the host machine.

Users of the M-Grid system are presented with a web page which allows them to either send a computing task to another computer or to volunteer their own computer to perform a task for others.

This can be used as a teaching tool and Walters *et al* have suggested three teaching scenarios

- To demonstrate to students a distributed system in action
- To explore algorithm distribution which enables them to understand the method of dividing up a problem for distribution. Students can experiment with different configurations without having to write any applets or code.
- The more able students can code distributed programs by writing Java applets and solving problems in a distributed way.

Wang *et al* (2002) have created a groupware application for teaching that enables the teacher to guide students step by step through an application and allowing them to annotate directly on the student's interface. In addition other tools such as messaging and chat were available. Their system implemented a tracking mechanism which stored user name, IP address, time stamp, chat messages in a database for future analysis. At this point no analysis had been attempted.

Bouras and Tsiatsos (2002) have sought to construct a collaborative e learning environment by basing it upon well established collaborative learning techniques. These include

- Brainstorming – using an audio collaboration function or text chat. The tutor can upload a question into the interface and learners can use a brainstorming tool to write and attach their ideas.
- Think pair share – using an audio collaboration function or text chat to present a question. Participants can then turn into whisper mode which allows them to talk with one other person only in secret. After the assigned time the tutor rings the bell or sends a text message to all participants who exit from whispering mode and share the results of their discussion.
- Jigsaw – The tutor sets up four separate session areas and allocates a number of participants to each area. The learners receive an automated message telling them which virtual room to go to and what their task is. Documents can be assigned to each area. The four areas correspond to four different parts of a bigger problem. After a predetermined time everyone comes back to share their solutions and the bigger problem can be solved.
- Quickwrites – The tutor constructs four virtual classrooms and presents the learners with a task. Learners move to their classrooms and discuss the problems. One person then writes up a summary of the group discussion and loads it up into the share space and in the main classroom the different results are shared.
- Structured academic controversy – Again virtual classrooms are used, the tutor uploads a topic with two different viewpoints. Learners are divided into two pairs, each pair needs to advocate one side of the argument and prepare a presentation. Each pair presents its position to the other pair, and then the learners debate and provide more evidence. Finally a vote is taken on the outcome.

Bouras and Tsiatsos (2002) conclude that in order for full collaborative working that mimics classroom, work an online environment may need to have some extensions including audio and visual capability as well as virtual classrooms and a private whispering mode.

Boyle *et al* (2003) have developed the QCDOC supercomputer which is designed for the highly specialised task of calculating results for lattice quantum chromo dynamic systems. This is an example of collaboration software between University laboratories to enable joint work to be done in a multi user environment. Increasingly collaboration is taking place online for a variety of work based tasks CSCW and for learning based instruction CSCL.

Brocks *et al* (2003) have developed a multi agent based approach called the MACIS framework which introduces collaborative elements in a natural way. This framework is constructed on four layers:

- The **user layer** which provides a representation of the current user in terms of the task that they are required to perform.

- The **user interface layer** this is compose of user interface agents (UI agents) which perform specific tasks and may form organisational relationships to complete more complex tasks
- A **service layer** this comprises those agents which do not provide a user interface
- The **persistence layer** which is responsible for the storage and retrieval of information.

Brocks *et al* have concluded that the multi agent approach is ideally suited to support virtual teams particularly in the realm of discussing documents.

Nick Jennings of Southampton University has written about a similar agent base virtual laboratory called Trilogy the purpose of which is primarily for the training of research students and also for the management of information and tools as well as to demonstrate the development of agents and virtual laboratories across three collaborating sites.

The virtual environment is assisted by agents (independent software entities that act inside the environment) which perform three tasks

- Personal assistants- performs the task given to it by the user but may also use intelligence in order to suggest documents that match the user's interests
- Mediators – they may suggest collaboration between users with common interests and point to information and provide tools
- Resource agent – providing access to resources through a common interface

## 4.6. Problems of CSCL

Valcke and Martens (2006) raise quality issues concerning CSCL environments and conclude three things

- More accurate research methods are required to obtain details about CSCL processes
- A higher validity of research methods is required. Content analysis has a weak theoretical basis.
- A higher reliability of research methods is required there is a lack of replication studies

Lipponen considers the results of empirical studies in CSCL research but concludes that "It is difficult to make any solid conclusions" due to the great variety of techniques and technologies used, purposes sought and applications applied.

He also examines the tools used for collaboration and concludes that currently almost any web based application is labelled collaborative, which is too loose. Most internet tools are not robust enough for classroom use. Basic internet chat, bulletin boards or email do not organise conversations well enough for learning and they are not designed for pedagogical purposes. These ideas are controversial and not all researchers would agree with them. Some would argue that ideas can be exchanged in all mentioned mediums and can be learned from. According to Lipponen he believes that CSCL needs specially designed programs for educational use which can build up collaborative knowledge. Technology does not solve the problems of CSCL but may act as "a Trojan mouse" and serve as a catalyst for change (subconsciously). He proposes that innovative pedagogical practices should take priority and technology brought in as a secondary item to make them work.

Dillenbourg (1999) expresses a similar view to Baker, Greenberg and Gutwin (2002) in the field of collaborative learning, and following his research program entitled "Learning in Humans and Machines" which gathered together twenty scholars from the disciplines of Psychology Education and Computer Science, they discovered that their group did not agree on any definition of collaborative learning. There is such a large variety of approaches to collaborative learning that it is impossible to provide a single definition. "When a word becomes fashionable – as it is the case with collaboration- it is often used abusively for more or less anything." Dillenbourg (1999)

Dillenbourg 1999 has highlighted issues with research methodology for measuring the effects of collaboration on the learning process. The first issue is "What effects are really being measured". Collaboration is difficult to measure in and of itself and so we measure instead some effects of collaboration such as how well a task is performed or how accurately a result is obtained. Most research attempts to measure effects through an individual, pre-test followed by an individual post-test and the difference is measured with respect to task performance, but these results have been criticised as being too qualitative leading to limited conclusions.

Dillenbourg's second issue concerns the method of evaluation. Collaborative learning is often assessed by measuring individual task performance but objection has been raised that a valid assessment would be required for group performance. Unless some way of measuring group performance becomes available then the existence of a hypothetical ability to collaborate remains to be established.

## 4.7. Review of CSCL Research Findings

Rahikainen *et al* (2001) have conducted a study of 21 ten year olds in a CSCL environment over four weeks. They studied the area of biology under the topic of adaptation. Students worked in groups of two or three and spent half of their time sharing knowledge using CSCL. Data was collected by two video cameras for later analysis. The work showed that whereas student levels of advancement was good with the more able students, there were difficulties with the less able students and some had still not learned anything by the end of the process. They concluded that teachers need better instructions in order to guide different levels of students. The implications for the SIM20 environment are that the less able students need careful monitoring.

Tapola *et al* (2001) have demonstrated that the main challenge of computer supported learning is that some students participate more than others. The results of their study indicate that whereas the more able students were actively engaged the least able were not. However, no gender differences were discovered. They conclude that students that have a problem being motivated may not do well with CSCL and may require greater tutor input.

Varey (1999) discusses her experience in remote teaching and evaluation of course work using Net Meeting. She claims her experience of collaboration as positive and includes the following benefits

- Increases flexibility of lecturer time
- Sharing of developmental costs
- Better curriculum
- Staff development
- Student enjoyment of using the technology
- Student enjoyment of involvement with other students

McCarthy and Boyd (2005) have remarked upon a new use for instant messaging and group chat which enhances the learning process within shared physical spaces. They comment upon the use of digital backchannels in the context of an academic conference where during a speaker's presentation chat channels are opened up and all participants can communicate using laptops thereby adding information to what is being presented. They also point out some disadvantages of that facility leading to distraction from the presentation itself.

Joiner *et al* (2006) have done a study of educational experiences using digital technology. They designed two learning experiences; the first had a goal for the students to accomplish while the other did not have a specific goal. They concluded that students overwhelmingly preferred the goal version to the non-goal version. They also point out that one of the reasons that digital games are so involving and motivating is the presence of a continuous series of ever increasingly difficult goals. This has a clear implication for the design of any collaborative on line system where the goals must be clearly defined in order to make the experience compelling. The design of any interface must therefore include consideration of goal setting, target achievement, and personal reward.

Kester *et al* (2006) have investigated the process of effective learning by examining the place of written supportive help and a schema explaining the process of solving the problem. They performed experiments where they presented the support and the schema either before or during the problem solving period and found the following conclusions

| | 1st Phase | 2nd Phase | Outcome |
|---|---|---|---|
| CASE 1 | Schema → | Problem Support Info | Higher learning Efficiency |
| CASE 2 | Support Info → | Problem Schema | Higher learning Efficiency |
| CASE 3 | Support Info → Schema | Problem | Lower learning Efficiency |
| CASE 4 | | Problem Support Info Schema | Lower learning Efficiency |

**Table 7**

The implications of this are that when the support information and the schema come together there is a lower learning efficiency. This is probably due to "temporal split attention" according to the findings of Kester *et al*. However there was no such temporal split if the support information or the schema were presented before the problem was tackled.

This suggested that any interface that is constructed to assist collaborative learning needs to ensure that supportive information and schematic information are presented at separate times.

Dillenbourg has done a number of tests to compare the difference in learning between pairs of students and individual students. In a number of cases he found that learning by pairs was less effective than learning by individuals. He interpreted this as a "split interaction effect" which suggested that pairs of learners would suffer from interference between the two interaction processes as well as the social interactions between the learners and the interactions with the material. This suggests that cognitive load can in some circumstances be greater in pairs than with individuals, leading to a loss of focus and learning. (Dillenbourg 2006) However, when using online animation as a delivery mechanism he found the opposite results. The animated pictures had a positive learning effect with pairs and not with individuals. Dillenbourg suggests that this finding is due to a lower cognitive load recorded by the pairs. The weakness of these two divergent results is clear and rests on the fact that there is no objective measure of cognitive load which might lead to different results. More work needs to be done in this area and this might be a suitable topic for further investigation in this research.

Graves and Klawe (1997) have built a CSCL environment which allows pairs of players to build a house together using such online activities as a graphical interface 2D and 3D views, sound feedback, real time written communication and real time spoken communication. They have found that in practical use with a 134 schoolchildren that academic results in the math area had increased significantly. In particular an interesting gender difference was discovered which revealed that males responded well to having a specific goal while females responded more to being able to speak to and see an image of their collaborator. There is also anecdotal evidence that some females were more interested in communicating than in performing the task. These gender activities should be used as guidelines in the design of learning activities. The conclusion points to further work being done to answer the question of whether enhancing the communication will improve the learning gain. It is suggested that one interesting approach would be to create an entirely user configurable learning environment so that they can choose what type of communication channels they want to use. This would help to define which elements of CSCL were valuable to different types of learners. This Meta design would lead to another level of research.

Klawe, M. (1999) has performed studies on learning through collaborative computer games. Her results demonstrate that games can be very effective in increasing motivation and achievement in learning. In particular major results have demonstrated that by placing two students to work together on a single computer has a positive effect on achievement and motivation. A further result concerns paired working of students where each student worked on their own computer and shared the same online environment in real time. A field study with a 134 students found that playing in the online environment led to significant learning gains and improvements in understanding were achieved after only a short period of 30 minutes. One of the most significant results was obtained by "requiring the achievement of a specific numeric goal versus a more open ended maximisation goal".

The implication of Klawe's work for the design of a collaborative interface for paired working of students is that the implementation of specific numeric goals will be more effective.

Lawless and Allan (2004) have considered the unwanted outcome of stress generated in collaborative e learning environments. They contend that stress can be designed out of an on line collaborative exercises by a careful management of the working processes. The follow Palmer in defining stress as "When the perceived pressure exceeds your perceived ability to cope". Since stress is a matter of perception one student may feel stressful while another may feel the situation is enjoyable. In order to minimise stress a number of steps need to be taken in designing the interface.

- Technology should be as user friendly and trouble free as possible
- The environment should be a comfortable and accessible learning space
- A clear specification to the students taking part of the minimum technology level required
- Clear specification to the students of the minimum prior knowledge required
- The running of induction courses of the interface system
- The maintenance of a help desk to address problems

Further individual stress can be reduced by

- Providing sufficient time for asynchronous working including time taken to build relationships
- Removing time pressure from the set goals
- Ensuring group roles are clearly defined
- Monitoring group cohesion

### 4.7.1. HCI Help Systems and effects of Learning

Bartholome *et al* (2005) have made a study of help effectiveness and the construction of well designed help systems. They have found that students accessing help materials at a high rate made more correct decisions and spent less time on wrong paths than those who did not. They also discovered that Students with low prior knowledge did not do as well as those with higher prior knowledge. Further research also showed that students with low work avoidance were equally successful as those with high work avoidance or in other words motivation had no discernible effect on help seeking and performance. Batholome's team conclude that help functions by themselves are not effective. Further work is needed in this area to see if this is borne out within our own e-lab or not and an inclusion of and monitoring of a help system will allow us to validate these claims.

Dillenbourg (2006) also comments on another interesting result concerning how pairs use help functions provided by the software. What they discovered was that one might expect that pairs working together would use the help system less frequently than a single person because they would be helping each other. However what was found was that pairs triggered the help system more frequently than individuals. This might be a result of disagreement and an appeal to higher authority or the sparking off of different ideas in the individuals creating more questions. Dillenbourg's conclusion is that we cannot predict how social interactions of pairs will affect individual cognitive processes. One therefore cannot generalise from individual learning to group learning. Consequently one will need to continue conducting experiments in both settings.

# 5. Computer Supported Collaborative Research (CSCR)

Handoko (2005) describes a newly developed on-line scientific web log which enables scientists around the world to perform an on-line collaboration over the internet. This can be said to be one of the first examples of what we would like to call CSCR (Computer Supported Collaborative Research) Hinze-Hoare (2006c)

Handoko indicates the following important characteristics of research collaboration.

- A mechanism for managing the knowledge generated in research collaboration is needed (this may involve a kind of database which allocates activities and content to different research entities).

- Equal treatment and non discrimination of participants – this will facilitate collaboration amongst multi institutions and multi nations (every participant receives the credit they deserve).
- Rapid research progress is often made over relatively short periods which often lead to tight competition between research groups and the race to publish and achieve primacy. Any tool which can facilitate on-line publication or preprints to a site such as arxiv.org automatically would be of benefit.

The system should be able to be maintained, used and contributed to by every member of the research group and by no one else. It is important to maintain a secure area where work is developed before it is published.

- The system will need to accommodate every aspect of scientific collaboration.
- The ability to share all kinds of data with members which may include sound video image text etc.
- A database for publications and member's details
- A balance sheet shared with all members to enable transparent management of funds.
- A common shared agenda to enable members to set collaborative schedules.
- An instant message system equipped with an alert to allow constant communication between participants.

Hinze-Hoare (2006c) has suggested differences between CSCW, CSCL and CSCR in terms of different working spaces. The primary difference between CSCR and CSCL is that a complete record of all interactions between participants is an important and necessary tool to evaluate the contributions of each member in a collaboration group which can later on determine "a fair capital share" if the undergoing research project is successful. Further differences will include the following

- **Knowledge Space**
  Research collaboration will generate its own knowledge base and a database system will be required which can store and retrieve this information as well as allocating ownership to individual contributions to ensure an appropriate apportionment of credit. It would be expected that this system would incorporate hypertext and links to bring cohesion to individual contributions.
- **Publication space**
  The publication of pre-prints and draft papers to online sites such as arxiv.org would be assisted by an automated process incorporated into the system.
- **Privacy Space**
  The research group will need a private area in which to work that is closed to non-group members. It is important to maintain a secure area where work is developed before it is published.
- **Public Space**
  The collaborative research group may wish to provide information upon the nature of the research which is being done, to encourage contributions, questions, raise issues etc. which can be place online in the public domain.
- **Negotiation Space**
  Group research may often introduce conflicts of opinion which need to be worked through. This is more difficult online and may involve intensive and protracted discussions.

Unlike CSCW and CSCL it has been shown that there is a case to be made for regarding CSCR as a separate and distinct area of investigation. Each of these domains has their own specification and requirements. The first two have much more than that including their own "conferences, journals and adherence" Stahl, G. (2003). The latter is yet to develop and is a potential fruitful area for future research. All three domains have commonality and dependency, and borrow from one another. However, CSCR has individual aspects which are not part of the other two, and consequently is distinct and should be treated as such.

| The Spaces required by each of the collaborative areas | | |
|---|---|---|
| **CSCW WorkingSpace** | CSCL LearningSpace | CSCR ResearchSpace |
| Communication | - | - |
| Scheduling | - | - |
| Sharing | - | - |
| Product | - | - |
| | Reflection | - |
| | Social | - |
| | Assessment | - |
| | Tutor | - |
| | Administration | - |
| | | Knowledge |
| | | Privacy |
| | | Public |
| | | Negotiation |
| | | Publication |

**Table 8**

# 6. Further Research Questions

- How do participants become aware of the benefits of collaboration via computer-supported tool, and how can these subsequently improve their learning?
- What kinds of strategies (collaborative strategies, self-regulated strategies, social interpersonal skills) do learners use in CSCL environment? and how much do they gain through the process?
- What theories of learning can be transferable to CSCL systems?
- What are the roles of teachers in the CSCL environments? What are their attitudes toward the CSCl systems? What makes them use or not use the systems? What kind of supports and training they need to integrate into their curricula?
- Does computer mediation require the development of new and special pedagogical techniques?
- How can best utilize the attributes of the CSCL systems in designing a particular subject domain? The best computer-supported tools should not simply offer the same content in a new format; rather they should provide new ways of thinking in those domains (Resnick, 1995).
- What are the important design considerations for developing CSCL applications? What are some of the problems of implementation? Koschmann,(1995)
- How to apply CSCW experience to CSCL? CSCW that supports business teams will not be the same for students in an educational setting. There is a need to redefine the role of individual, her responsibilities, the level of interaction, and environment (Olson et. al, 1993).

- How to marry methodologies from CSCW and educational research to CSCL? Webb (1993) identified that questionnaire and content analysis based on critical thinking and social interaction are powerful methods to study on-campus on-line conferencing.

# 7. References

## INDEX OF RESEARCHED REFERENCES